# Initial Offset Placement in p2p Live Streaming Systems

Chunxi Li and Changjia Chen
School of Electronics and Information Engineering, Beijing Jiaotong University, 3 Shangyuancun, Haidian District, Beijing 100044, P. R. China
cxl@telecom.njtu.edu.cn and changjiachen@ sina.com).

***Abstract*—Initial offset placement in p2p streaming systems is studied in this paper. Proportional placement (PP) scheme is proposed. In this scheme, peer places the initial offset as the offset reported by other reference peer with a shift proportional to the buffer width or offset lag of this reference peer. This will introduce a stable placement that supports larger buffer width for peers and small buffer width for tracker. Real deployed placement method in PPLive is studied through measurement. It shows that, instead of based on offset lag, the placement is based on buffer width of the reference peer to facilitate the initial chunk fetching. We will prove that, such a PP scheme may not be stable under arbitrary buffer occupation in the reference peer. The required average buffer width then is derived. A simple good peer selection mechanism to check the buffer occupation of reference peer is proposed for a stable PP scheme based on buffer width.**

## I. INTRODUCTION

During last several years, p2p data-driven live streaming [1] is one of the hottest applications, which attracted large number of users in a rather short period since appeared [2-9]. Abundant program resources, convenient access method, low configuration requirements, good playback quality, free service, they all contribute to the great success of these systems.

There are two types of implementation for p2p live streaming systems: small buffer approach and large buffer approach. In the small buffer approach, peer uses a buffer with a small buffer width. For example, the buffer width is fixed to 128 chunks in PPStream. The playback duration of contents in a full loaded buffer is about 22 seconds. In other sides, the buffer width in PPLive is about 2000 chunks and the playback duration of contents in a full loaded buffer is about 210 seconds. The advantage for small buffer approach is fewer overheads in buffer message exchange and smaller playback delay. However, the scalability will be a big problem when tens of thousands of peers share a channel. Apparently, the drawback of large buffer approach is the large playback delay since all chunks in the buffer of a peer are already sent out by seed but not played by the peer. However, playback delay is not very sensitive for TV watching only. Except to the larger playback delay, large buffer approach is out performance small buffer approach in almost all other aspects. For examples, the large buffer approach is good in its scalability. Furthermore, its startup latency (the time when peer registers himself to a channel to the time when the playback starts in his screen) is shorter than the latency of small buffer approach in general. A newly joined peer to a large buffer system could fetch chunks that are already fetched by all online peers. His download rate at this stage can be much higher than the playback rate. However, in a small buffer system, the environment in buffer fetching is the same for stable peers and newly joined peers. Their download rates will be constrained to the playback rate in consequence of the chunk availability. Based on the same reasons, large buffer fading can be quickly recovered in a large buffer system. In this paper, we will study the startup process in the large buffer approach of p2p live streaming systems.

In a small buffer system, every peer is busy in fill his buffer by fetching chunks from other peers. Since only fixed number of few chunks is opened for share, they must be competed by all peers. A newly joined peer could not do any better than that of other peers. There is no room to design a new protocol for a newly joined peer [10-12]. However, in a large buffer system, large number of chunks is opened for share. Buffer of every online peer is largely filled from its head. Peers already online are competing (exchange) chunks in their buffer tail. A newly joined peer could choose those chunks that already fetched by almost all other peers to build his empty buffer. Thus, the initial download rate will be much higher than that of other peers since there is no competition for those chunks. In other words, there would be a startup stage for a newly joined peer in a large buffer system. In this stage, a newly joined peer acts differently from what when peer goes stable. By our knowledge, large buffer approach has not been studied before by any published literatures. In this paper, we will study how a newly joined peer chooses his first chunk to fetch as his buffer head. We will name the ID of this chunk as the initial offset, and name this problem as the initial offset placement.

Higher initial download rate is only one aspect need be considered in the initial offset placement. In fact, all afterward share provided by a peer is determined by his initial offset. Assume the initial offset is $\theta$ and the buffer starts to be drained at time $t_{off}$ for a peer $p$, then all chunks with ID larger than $f_p(t)=r(t-t_{off})+\theta$ fetched by peer $p$ will be opened for share at time $t$. Where $r$ is the playback rate of the video. We will assume a CBR video in this paper for simplifying the discussions. If we assume tracker (seed) feeds chunks with service curve $s(t)=rt$, then the buffer space that the maximum number of buffered chunks by peer $p$ is $L_p=s(t)-f_p(t)=rt_{off}-\theta$. In the protocol design of a large buffer system, a placement process has to ensure the resulted offset lag $L_p$ is large and similar for all peers.

Since a tracker in a p2p live streaming system has to serve many channels (hundreds in general), we certainly do not wish the tracker also buffers large number of chunks for each channel. Therefore, the buffer width of a tracker is significantly smaller than the buffer space of peers in a large buffer system. There is a problem that, when a peer joins an empty system without any other online peers, his initial offset must be a chunk in the buffer of the tracker currently. The only way to have large buffer space in this case is to use a large offset setup time $t_{off}$. In other sides, when a peer joins a crowd system, he is better to choose his initial offset that is saved by all other peers. This chunk is in general already discarded by the tracker in a large buffer system. Therefore, we have to shorten the offset setup time $t_{off}$ for getting same buffer space. It is not a good protocol design to detect the number of online peers and adjust the offset setup time accordingly for a newly joined peer in a practical system. A protocol named as *proportional placement* (PP) scheme is proposed in this paper to overcome this problem. In this scheme, a constant offset setup time $t_{off}$ is adopted. A newly joined peer places his initial offset $\theta=f_q(t_0)+\alpha L_q(t_0)$ according the offset $f_q(t_0)$ and buffer space $L_q(t_0)$ reported by the first report of his first neighbor $q$ at time $t_0$. The initial offset is the offset advanced by a proportion of the buffer space of the first neighbor. The proportional coefficient $\alpha$ is a constant. We will show that, with this simple scheme, buffer spaces of peers will automatically converge to a designed stable value from any given tracker buffer width.

Above discussed PP scheme is specified by the buffer space $L_p=s(t)-f_p(t)$ of a peer. Buffer space is also called as the offset lag [2-6] in some paper. So, we will name above scheme as the PP scheme based on offset lag. Offset lag of a peer is independent to any chunk fetching strategy and chunk fetching result. Furthermore, in a CBR system, offset lag of a peer is also independent to the time. In a real system, a peer will adjust his buffer width with the most advanced chunk he knows currently. In other words, the buffer tail is always occupied by a fetched chunk. Thus the offset lag, the number of chunks a peer really buffered and the current buffer length of a peer are different in a real system. Above PP scheme is not practicable in a real system because it is based on both the offset lag and the buffer message of the first neighbor. Peer only reports his buffer width to other peers in a real system. To find offset lag, a peer has to know the service curve from the tracker. However, messages from a peer and from a tracker are received at different time and have different report periods. It is easier to place initial offset only based on the buffer width of neighbor peers. Unfortunately, PP scheme based on buffer width is not stable in general. If many newly joined peers place their initial offset based on the buffer width of poor neighbors (with a small buffer width), offset lags could be unbounded. Some selection mechanism in reference neighbor is inevitable to ensure the PP scheme places initial offset based on the buffer width of a good neighbor peer. A simple good peer selection protocol will be discussed in this paper.

Facility the initial chunk fetching is the other reason to place initial offset based on the buffer width of neighbor peers. For an extreme example, if the offset lags of all neighbor peers are very large but the buffer widths are small. PP scheme based on offset lag may place an initial offset larger than buffer widths of all neighbor peers. In this case, the newly joined peer has to fetch chunks from either the tracker or waiting his neighbors to fetch those chunks first. The initial download rate is largely limited. However, the initial offset when placed based on buffer width is always inside the buffer of one neighbor peer. With large probability, it is also inside the buffers of other neighbor peers if a good fetching strategy is adopted by the system. In this case, the initial download rate for a newly joined peer will be high. Above discussion indicates that, the fetching strategy adopted by a system will affect how the initial offset is placed. If buffer widths of stable peers are small in large probability, the offset lags resulted from PP scheme based on buffer width would be unbounded. A PP scheme is called stable if it results bounded offset lags. We will show that there is a requirement to the average buffer width on stable peers for a stable PP scheme based on buffer width. Therefore, this average buffer width should be enforced by the fetching strategy of a system, when PP scheme based on buffer width is adopted in this system.

This paper is not only an analytical study to a theoretical model. In fact, the work in this paper is motivated from our measurement on PPLive, one of the most popular IPTV applications to date. Our measurement shows that PP scheme based on buffer offset is very likely adopted by PPLive in its peer startup process. In other words, PP scheme based on buffer width is a good approximation to the initial offset placement for peers in the PPLive. The values of parameters for this scheme used in PPLive are inferred. The stability condition is checked as well.

**Related Works:**

General architecture and initial experiment of a p2p live streaming system is discussed in [1]. After that, many works of measurement-based studies on p2p live streaming video are published recently, for example [2-9]. The concept of buffer progress and related parameters such as buffer width, playable video and peer offset used in this paper are defined in [2-6]. Most of those works are descriptive. They describe how such a system is working and how to measure such a system. Statistics are collected and analyzed to reveal how peers are cooperated mainly in their stable states. Instead, in this paper we will try to take a microscope view, to study how a peer starts his corporation at very beginning.

Protocol studies of p2p streaming systems in literatures are almost all for the small buffer approach [1, 10, 11, 12]. Hence, the fetching strategy at an equilibrium environment is their main studying object. Same strategy is adopted by stable peers as well as by a newly joined peer. Startup performance is evaluated in [12] based on this single strategy approach. A newly joined peer could not do better than that of any stable peers.

By our judgment, the difference between small buffer approach and large buffer approach is first noticed by [6]. A tiering effect in the playback lags of PPLive is discovered in [6]. However, authors of that paper try to interpret this effect by a "tree like structures" inside a mesh-pull p2p network. This interpretation is true only for small buffer approach since in this kind of system, the length of peer buffer is small and fixed. Based on our measurement, the PPLive adopts a variable buffer length with very large buffer space (more than ten times of the buffer space of small buffer system). The largest chunk ID in the buffer of each peer is very close each other and independent to offset lags. Therefore, the "tree like structures" may not be a good interpretation to the tiered offset lags in a large buffer approach. If the "tree like structures" is not the main reason for the tiering effect, a natural question is to ask if the initial offset in a large buffer system is placed by certain specified mechanism. The tiering effect is made up by this mechanism.

Studies on the content distribution in CDN environment mixed with servers and peers will help us to answer above question. It is shown by [13-16] that, there is a phase-transition point $C(t)$ for each time $t$ in a mixed distribution environment, any chunks with ID less than point $C(t)$ will be fetched readily by any peers. In a small buffer system, the phase-transition point is in the head of shared buffer space. A newly joined peer has to compete chunks above $C(t)$ with other peers at his startup stage. However, in a large buffer system, the phase-transition point is in the tail part of shared buffer space. A newly joined peer could chose his buffer head (initial offset) below $C(t)$ to avoid the competitions with other peers at his startup stage.

The organization of the paper is as follows. In Section 2, we will discuss possible models of initial offset placement in p2p live streaming systems after a briefly outline on p2p live streaming systems. The *proportional placement* (PP) scheme is emphasized in this section. This scheme is verified in section 3 through a real p2p live streaming system PPLive. The stability of a PP scheme based on buffer width and the average peer buffer width for a stable placement is studied in section 4. Section 5 concludes the paper. Startup Process and Initial Offset Placement

## II. Startup Process and Initial Offset Placement

### *2.1 A Brief on p2p Live Streaming System*

A p2p live streaming system uses few servers (named as tracker) to support large number of audiences (named as peer). In a typical p2p live streaming system, the video data is divided to chunks identified by continuously assigned sequence numbers. The sequence number is also called the chunk ID in many papers. Content server (or Seeder) of the system injects chunks one by one to the system (actually it is selected or asked by peers). Every peer in the system has a buffer organized by chunks. A buffer message (BM) is an abstract description of this buffer. BM consists of an *offset*, which is the ID of the chunk at the buffer head, and a sequence of {0,1} in other words a bit map of the buffer to indicate which chunks are stored in this buffer. A value of 1 (0) at the $i^{th}$ position indicates that the chunk with an ID $offset + i-1$ has been (has not been) stored in that buffer.

The startup process when peer joins a p2p live streaming system is studied in this paper. We will use the *host* to name a newly joined peer. The name *peer* will be left to address other peers the *host* connected. When a host joins, he first registers himself to a tracker. A list of current online peers is returned from the tracker after this registration. The host will try to connect peers in the list then. A peer will send his current buffer message to the host once he is connected. The host then will choose a chunk ID in the streaming as a *start point* of his buffer to fetch chunks above this ID from other peers. This start point is also named as the *initial offset* in this paper. After certain time, the host will drain his buffer with the playback rate.

We will name the peer as the *first peer* of a given host if he returns his buffer message to this host earlier than all other peers do. In the following subsection, we will discuss a simple model on the placement of the buffer start point. In this model, the placement is mainly based on the buffer message from the first peer of a host.

### *2.2 Initial Offset Placement Based on Offset Lag*

In this subsection, we will provide a very simple model on the initial offset selection process in a live streaming system. Someone may criticize this is over simplified to a real system. However, it is a useful start to comprehend basic considerations in the more complex real system. In this model we will use $s(t)$ to indicate the *service curve* of the media stream. Where $s(t)$ is the largest chunk ID in the system at time $t$. For simplicity, we assume the media stream is CBR with a constant playback rate $r$. Peer drains his buffer with this rate. The chunk ID in the buffer head of a peer $p$ at time $t$ is $f_p(t)$. We will name $f_p(t)$ as the *offset curve* of this peer. Under the CBR assumption, the difference between service curve and the offset curve of a given peer is independent to time. We will name the difference $L_p=s(t)-f_p(t)$ as the *offset lag* of peer $p$. It is also the *buffer width* of peer $p$ if the buffer tail is always reserved for the most advanced chunk currently. Furthermore, we assume the tracker (or seeder) buffers the media contents with a width $W_{tk}$. Thus the tracker has a offset curve $f_{tk}(t) = s(t)-W_{tk}$.

Now let's consider the startup process of a host $h$. Assume at time $t_0$ the host chooses a chunk ID $\theta$ as a *start point* for his buffer head, and then fetches chunks to fill his buffer. After a time interval $\tau_s$, the host starts to drain his buffer at time $t_0+\tau_s$ with the playback rate [8]. The offset curve of the host then reads

$$\begin{cases} f_h(t) = r(t - t_0 - \tau_s) + \theta \\ L_h = s(t) - f_h(t) = s(t_0) - \theta + r\tau_s \end{cases}, \text{ for } t \geq t_0 + \tau_s \quad (1)$$

For minimize the load of tracker, a system designer certainly wishes chunks are fetched from other peers instead of from the tracker, so $t_0$ should be chosen as the time that at least one peer $p$ has returned his buffer message. The initial offset $\theta$ should be larger than the offset $f_p(t_0)$ of this peer. In other side, big diversity in offset lags are no good for content sharing among peers, so a system designer will wish to control the difference between the lags of the host and other peers:

$$L_h - L_p = f_p(t_0) + r\tau_s - \theta \quad (2)$$

One may think to make the left side of (2) to be zero is a good criterion in selecting the start point $\theta$. We will name this placement scheme as *fixed padding* (FP) since the start point $\theta$ is simply the offset reported by the first peer with a constant padding $d=r\tau_s$:

$$\theta = f_p(t_0) + r\tau_s = f_p(t_0) + d. \quad (3)$$

There is not much design freedom in FP scheme. When a host joins an empty system without any other peers, he can only fetch chunks from the tracker, thus we will have $L_h \leq W_{tk}$ after padding in this scheme. One can easily find that then all peers in the system will have the same offset lag $L \leq W_{tk}$ under this placement scheme.

Buffer width is an important design parameter on the playback performance. Larger buffer width in peers will result better playback continuity since larger buffer fading in one peer can be quickly recovered from his neighbors. In other sides, large buffer width is definitely unwished to a tracker since a tracker has to host many channels. FP scheme is not good in practice since it cannot fulfill those two goals of larger buffer width in peers and smaller buffer width in tracker at same time.

Next let's consider a more practical scheme as *proportional placement* (PP). We are such name this scheme because the start point $\theta$ is the offset of the first peer with an advance that is proportional to the offset lag reported by this first peer with a constant *placement coefficient* $\alpha<1$:

$$\theta = f_p(t_0) + \alpha L_p. \quad (4)$$

In this scheme, the offset lag $L_h$ of the host is

$$L_h = (1-\alpha) L_p + r\tau_s. \quad (5)$$

Since the first peer of current host is a host when he joins the system, we can put (5) as a more familiar mathematical formula $x_{n+1}=bx_n+c$. Obviously it is a contraction mapping when Lipschitz condition holds for $b<1$. One can easily concludes that (5) is also a contraction mapping with $L^*=r\tau_s/\alpha$ as its stable point, which is independent to any initial value.

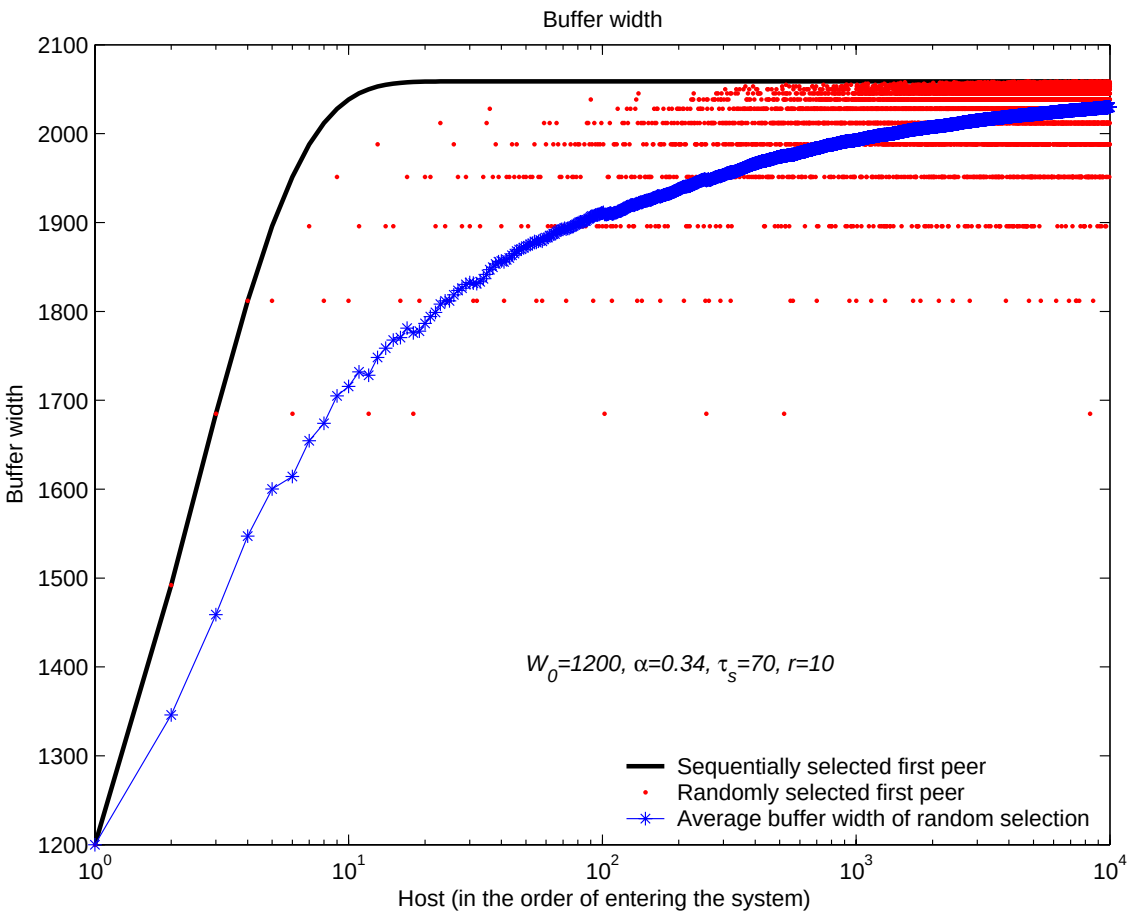

Figure 1. Offset Lag in PP scheme with different first peer selection methods

Self-stabilizing is the most attractive property of PP scheme. This self-stabilizing can be checked by following equations.

$$\begin{cases} L_h - L^* = (1-\alpha)\left(L_p - L^*\right), & (6.1) \\ L_h - L_p = \alpha\left(L^* - L_p\right), & (6.2) \end{cases}$$

Equation 6.1 shows that the offset lag of a host is always closer to the stable point $W^*$ than the offset lag of his first peer. Equation 6.2 shows that, the offset lag of a host is larger than the offset lag of his first peer if the offset lag of first peer is less than the stable point $L^*$. Otherwise, the offset lag of a host will be smaller than the offset lag of his first peer if the offset lag of first peer is larger than the stable point $L^*$. In this way, a system can accommodate two designed buffer widths: a smaller track buffer width $W_{tk}$ and a larger peer buffer width $L^*$. Peer will have a buffer width close to the tracker buffer width when there are few peers in the system. Along with many peers join the system, the peer buffer widths will converge to the designed stable point $L^*$. Figure 1 shows host buffer width under PP scheme when the first peer is selected sequentially or randomly. We say the first peer is selected sequentially if the first peer of each host is the peer that joins the system just before this host. We say the first peer is selected randomly if the first peer of each host is random one among those peers that join the system before this host. In a real system, the first peer is selected neither sequentially nor pure randomly, hence the host buffer width will fall in between of those two curves in this figure.

### *2.3 The PP Scheme Based on Buffer Width*

The PP scheme based on the offset lag is stable with a limiting lag $L^*=r\tau_s/\alpha$. However, there is a problem on the initial chunk fetching. In a real system, a first peer may only have chunks with ID below the initial offset placed by above PP scheme. For example, a first peer has an offset lag of $L_p$=1000 but only has 50 chunks in his buffer. If a host uses 0.3 as his placement coefficient, then the initial offset $\theta_h=f_p(t_0)+300$. In this case, the host cannot find any chunks above $\theta_h$ from this first peer. In practice, we will name the buffer width as the difference between the newest (largest) and oldest (smallest) chunk ID in a buffer message advertised by this peer. The buffer width of a peer is no longer a constant even under CBR assumption. In this paper, we will denote by $W_p(t)$ the buffer width of a peer $p$ at time $t$. Instead of offset lag, a host can use the buffer width of his first peer to place the initial offset. We will name this placement scheme as the PP scheme based on buffer width. The initial offset placed by PP scheme based on buffer width will be

$$\theta=f_p(t_0)+\alpha W_p(t_0). \tag{7}$$

When PP scheme based on buffer width is used, the chunk corresponding to the initial offset is always inside the buffer range of the first peer.

*2.3.2 Stable Condition in PP Scheme Based on Buffer Width*

While the PP scheme based on the offset lag is stable with a limiting lag $L^*=r\tau_s/\alpha$, the PP scheme based on buffer width may introduce an unstable placement. In other words, let $f_1(t), \ldots, f_n(t), \ldots$ be offset placed for sequentially joined peers $p_1, \ldots, p_n, \ldots$, the PP scheme based on buffer width can not guarantee bounded offset lags $L_n=s(t)-f_n(t)$ as $n\to\infty$.

To study this problem, let $t_h$ be the time when the first peer reports his buffer message. For simplicity, we will call $t_h$ as the *host up time*. Then the offset setup time $\tau_h$ is the time interval from the host up time $t_h$ to the time $t_{off}$ when the host starts to drain his buffer.

$$\tau_h\approx t_{off}-t_h \tag{8}$$

The offset of the host can be expressed as

$$f_h(t)=r(t-t_h-\tau_h)+\theta_h, \text{ for } t\ge t_h-\tau_h. \tag{9}$$

Now let's study a system with peers $\{1, \ldots, n, \ldots\}$ according their joining order. We will use $I(n)$ to denote the first peer of peer $n$. From (9), we have

$$\begin{aligned} f_n(t)&=r(t-t_n-\tau_n)+\theta_n \\ &=r(t-t_n-\tau_n)+f_{I(n)}(t_n)+\alpha W_{I(n)}(t_n) \\ &=f_{I(n)}(t)+\alpha W_{I(n)}(t_n)-r\tau_n. \end{aligned} \tag{10}$$

We define the scope factor $\beta_p(t)$ for a peer $p$ as the ratio of his buffer width to his offset lag:

$$\beta_p(t)=W_p(t)/L_p. \tag{11}$$

Obviously, the scope factor $\beta_p(t)$ is between 0 and 1 at any time. Compared with buffer width, offset lag is a more important system parameter. We will write equation (11) to an offset lag equation.

$$\begin{aligned} L_n(t)-L_{I(n)}(t)&= s(t)-f_n(t)-(s(t)-f_{I(n)}(t)) \\ &=r\tau_n-\alpha W_{I(n)}(t_n)= r\tau_n-\alpha\beta_{I(n)}(t_n)L_{I(n)}(t_n). \end{aligned} \tag{12}$$

After certain simplification, we will have an equation similar to equation (5) we have discussed before:

$$L_n(t)=(1-\alpha\beta_{I(n)}(t_n))L_{I(n)}(t_n)+r\tau_n. \tag{13}$$

Let's use $I^m(n)$ to indicate the $m^{\text{th}}$ first peer in the selection chain $P=\{n=I^0(n), I^1(n), \ldots, I^{K(n)}(n)\}$ and $K(n)$ is the length of this chain (precisely $I^{K(n)}(n)=1$). Let $\{t_n, t_{I(n)}, \cdots, t_{I^{K(n)}(n)}\}$ be the host up time and $\{\tau_n, \tau_{I(n)}, \cdots, \tau_{I^{K(n)}(n)}\}$ be the offset setup time of those peers in this chain. We will have:

$$\begin{aligned} L_n(t) &= (1-\alpha\beta_{I(n)}(t_n))L_{I(n)}(t_n)+r\tau_n \\ &= (1-\alpha\beta_{I(n)}(t_n))(1-\alpha\beta_{I^2(n)}(t_{I(n)}))L_{I^2(n)}(t_{I(n)}) \\ &\quad + r\left(\tau_n+(1-\alpha\beta_{I(n)}(t_n))\tau_{I(n)}\right) \\ &= W_{tk}\prod_{i=1}^{K(n)}(1-\alpha\beta_{I^i(n)}(t_{I^{i-1}(n)})) \\ &\quad + r\sum_{i=1}^{K(n)}\tau_{I^{i-1}(n)}\prod_{j=1}^{i-1}(1-\alpha\beta_{I^j(n)}(t_{I^{j-1}(n)})) \end{aligned} \tag{14}$$

For simplifying the discussion, we first assume the offset setup time is a constant $\tau_s$ for all peers and the chain length is infinity. If we write $\beta_{I^i(n)}(t_{I^{i-1}(n)})$ as $\beta_i$, then equation (14) is simplified to a limiting form:

$$L^* = W_{tk}\prod_{i=1}^{\infty}(1-\alpha\beta_i)+r\tau_s\sum_{i=1}^{\infty}\prod_{j=1}^{i-1}(1-\alpha\beta_j) \tag{15}$$

Equation (15) tells us that the limiting offset lag $L^*$ could be unbounded in a PP scheme based on peer buffer width. For example, let $\beta_i=1/i$, we will have following well-known limitation:

$$\prod_{i=1}^{\infty}(1-\alpha\beta_i)=\prod_{i=1}^{\infty}(1-\alpha/i)=e^{-\alpha}. \tag{16}$$

This will result an unbounded offset lag $L^*=\infty$. Following lemma will give a sufficient condition for a bonded offset lag:

***Lemma1:*** For PP scheme based on buffer width, the limiting offset lag $L^*$ is always lower bonded by $L^*\ge r\tau_s/\alpha$. Furthermore, if the scope factors $\{\beta_i\}$ are uniformly lower bonded above zero, the limiting offset lag $L^*$ is also upper bonded. More precisely, if there is a constant $\omega>0$ such that

$$\beta_i\ge\omega, \forall\, i \tag{17}$$

Then we have

$$r\tau_s/\alpha\le L^*\le r\tau_s/\alpha\omega \tag{18}$$

**Proof:** Since $1-\alpha \le 1-\alpha\beta_i \le 1-\alpha\omega < 1$, one has $r\tau_s/\alpha=r\tau_s\sum_{i=0}^{\infty}(1-\alpha)^i \le L^* \le r\tau_s\sum_{i=0}^{\infty}(1-\alpha\omega)^i=r\tau_s/\alpha\omega$ .

In general, a peer with large buffer width also has large scope factor. Therefore, if the first peer has small buffer width, a host could wait another peer with larger buffer width. We will call this as a good peer selection process.

This process will find one reference peer among initially connected peers. The placement scheme will place the initial offset based on the offset and the buffer width of this reference peer. We summarize this as following observation:

***Observation 1:*** For a p2p live streaming system, if the PP scheme is implemented based on peer buffer width, a good peer selection mechanism is better also implemented in the system. The good peer selection mechanism may simply check the peer buffer width $W_p$.

The stability in a buffer width based placement cannot be ensured by good peer selection only. In fact, how a host places his initial offset will not affect his buffer occupation status since a good buffer occupation is more related to the fetching strategy. To study offset lag more precisely must involve the study in buffer occupations. We will put this into our future research. In the following subsection, we will try to find the average buffer occupation level, which is required by a stable PP scheme based on buffer width.

*2.3.2 Average Buffer Width in a Stable PP Scheme Based on Buffer Width*

We can rewrite equation (10) to following equivalent:

$$\begin{aligned} L_1(t)-L_n(t) &= \left(s(t)-f_1(t)\right)-\left(s(t)-f_n(t)\right) \\ &= \alpha\sum_{i=1}^{K(n)} W_{I^i(n)}(t_{I^{i-1}(n)}) - r\sum_{i=1}^{K(n)}\tau_{I^{i-1}(n)} \end{aligned} \qquad (19)$$

For a given selection chain $P$, let's define the lag difference $\Delta L_p$ on this chain as the difference between the lags of the first and last peers in this chain:

$$\Delta L_p = L_{|P|}(t) - L_1(t). \qquad (20)$$

A stable placement needs lag difference $\Delta L_p$ on all selection chain $P$ been uniformly bounded. Based on this definition, equation (19) will have following expression:

$$\tfrac{1}{|P|}\sum_{p\in P} W_{I(p)}(t_p) = \tfrac{r}{\alpha}\tfrac{1}{|P|}\sum_{p\in P}\tau_p - \tfrac{\Delta L_P}{|P|},\ \ 1\in P \qquad (21)$$

To simplifying the notations, We will define the peer buffer width $E\{W|P, t_P\}$ sampled at peer uptime $t_P$ and averaged on $I(P)$ as

$$E\{W \mid P,t_p\} = \tfrac{1}{|P|}\sum_{p\in P} W_{I(p)}(t_p) \qquad (22)$$

Similar, we will define the averaged offset setup time $E\{\tau|P\}$ as:

$$E\{\tau \mid P\} = \tfrac{1}{|P|}\sum_{p\in P}\tau_p \qquad (23)$$

With these notations, we can use following lemma to summary above discussions:

***Lemma 2:*** On any given selection chain $P$, the averaged peer buffer width is approximately equal to the averaged offset setup time on $P$ scaled by a factor of $r/\alpha$:

$$E\{W \mid P,t_p\} = \tfrac{r}{\alpha}E\{\tau \mid P\} + O\left(\tfrac{1}{|P|}\right) \qquad (24)$$

If we assume the offset setup time $\{\tau_n\}$ are approximately iid random variables and uniformly bounded. By central limiting theorem [17, p345, theorem 9.5], we have a limiting form for the offset setup time:

$$E\{\tau \mid P\} = \tfrac{1}{|P|}\sum_{p\in P}\tau_p \to \tau_s,\ \ |p| \to \infty \qquad (25)$$

Where $\tau_s$ is a normal distributed random variable.

Furthermore, we assume the buffer width for any given peer is a stationary process. Homogenous chunk fetching strategy usually results stationary buffer occupation in a system [12]. Event though we do not discuss this issue in this paper, but this is a good approximation to the real system after peer goes stable. The time for peer goes to stable is relatively short in a p2p live streaming system, the selected peer will be in his stable range if the inter arriving time of peers is not too small. Under this assumption, we can omit the time instances in equation (23) and write the averaged peer buffer width with only the chain $P$.

$$E\{W|P\} = E\{W|P, t_P\}. \qquad (26)$$

Thus, we have proved following lemma:

***Lemma 3:*** The peer buffer width averaged on any selection chain with an infinite length is the same in probability sense:

$$E\{W|P\} = r\tau_s/\alpha,\ |P| \to \infty \qquad (27)$$

Not all mechanisms in peer selection will result infinite length of selecting chains. For example, if a mechanism always chooses peer 1 as the first (or good) peer for any host, the resulted chain length will be one. In practical system, the peer selection is mainly affected by the peer list returned from the tracker. In general, the members in a peer list are randomly selected by the tracker. Thus, a random selection mechanism in the first peer or good peer is a good approximation to a real system. For a random selection mechanism, we can show that

$$\Pr(|P|<M/\max(P)\to\infty)=0,\ \text{for } \forall M \qquad (28)$$

Based on above discussions we have following observation on the average of buffer width for stable PP scheme based on buffer width:

***Observation 2:*** For a stable PP scheme based on buffer width, the average peer buffer width $E(W)$ should be maintained at a level of $r\tau_s/\alpha$ by any fetching strategy:

$$E(W) = r\tau_s/\alpha. \qquad (29)$$

In PPLive we have measured a value of 208.3 for $E(W)/r$. With a placement coefficient of 0.34, the offset setup time is about 70.82s, very close to the offset setup time we have measured. Hence, the placement scheme used in PPLive is stable.

*2.4 Mapping the Simple Model into PPLive, a Real System*

PPLive is currently one of the most popular IPTV to date [2-6, 8-9]. In this paper, we will study the startup process of this system through data collected by our measurement. We will try to compare our simple models to the way a host places his initial offset on this real system. In the next section, we will verify that a very similar PP scheme is adopted by PPLive.

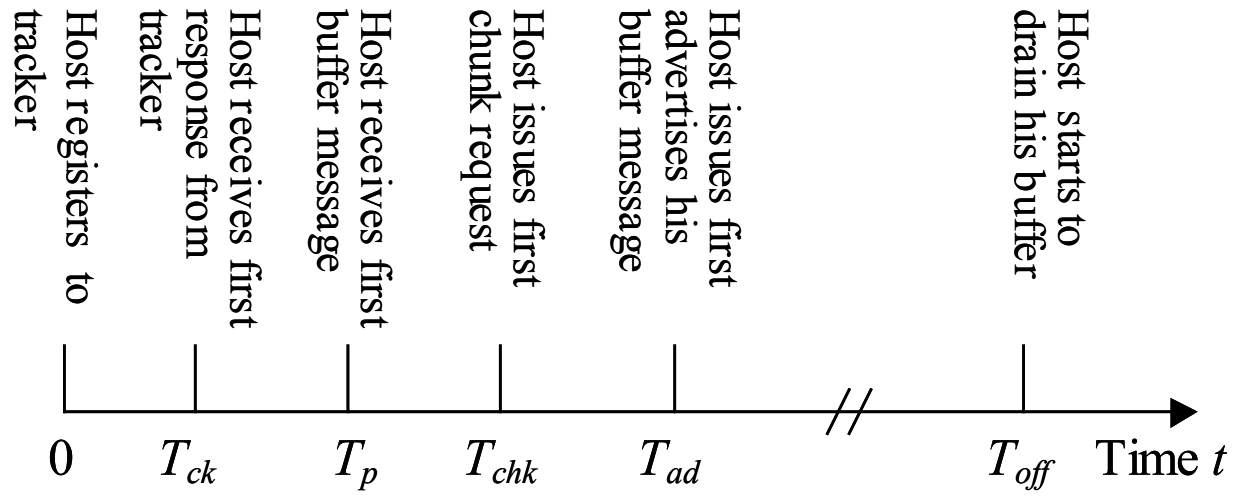

Figure 2. Events and their time in PPLive

Like all current p2p live streaming applications, PPLive is a proprietary system. We have to measure typical events and the time of those evens to guess the protocol used in this system. Those events and their time are shown in figure 2. The first event is the host registration, as we have mentioned before, a host must first registers himself to a tracker when he joins this system. We will take this time instance as our time reference 0. The host will receive initial response from the tracker after $T_{tk}$ seconds. We will name $T_{tk}$ as the *tracker response time*. Tracker response includes a list of peer addresses and two offsets *TkOffMin* and *TkOffMax*. The host will try to connect peers in the list then. The connected peer will send his current buffer message to the host. We will define the *peer response time* $T_p$ as the time when the host receives the first buffer message from the first peer. The host then will choose a chunk ID in the streaming as a start point of his buffer to fetch chunks above this ID from other peers. We will define the *chunk request time* $T_{chk}$ as the time when the host issues his first chunk request. The time $t_0$ (in subsection 2.2 that the host places offset start point $\theta$) is about the peer response time $T_p$. After certain time, the host will drain his buffer with the playback rate. We will name this time as *offset initial time* $T_{off}$ since the offset will progress continually after this time. The time interval $\tau_s$ in subsection 2.2 for offset setup is now

$$\tau_s = T_{off} - T_p. \tag{30}$$

Our measurement shows that the mean for $T_{tk}$, $T_p$ and $T_{chk}$ are relatively small with values 0.058s, 1.419s and 2.566s respectively. However, the offset initial time $\tau_s$ is relatively large and has a nearly constant value about 70 seconds.

Actually, a host will contact with the tracker and receive *TkOffMin* and *TkOffMax* roughly periodically at discrete time instances $\{t_i\}$. The value of $TkOffMax(t_i)$ at these discrete time instances is a sampling of the service curve $s(t)$ defined in subsection 2.2. Similar, the value of $TkOffMin(t_i)$ is a sampling of the offset curve $f_{tk}(t)$ of the tracker at time $t_i$. The difference of them is the buffer width $W_{tk}(t_i) = TkOffMax(t_i) - TkOffMin(t_i)$ of the tracker. Furthermore, we can show that the buffer width of the tracker is the scaled current playback rate $r(t_i)$:

$$r(t_i) = W_{tk}(t_i)/120. \tag{31}$$

In other words, a tracker will buffer 2 minutes contents for each channel in PPLive.

For this paper, we will adopt the definition in [6] to define the *buffer width* of a peer as the difference between the newest (largest) and oldest (smallest) chunk ID in a buffer message advertised by this peer. As we have defined before, the oldest (smallest) chunk ID in the buffer message is the offset of this peer. We also name the newest (largest) chunk ID in the buffer message as the *scope* of this peer since this is the most advanced chunk this peer ever fetched currently. In a real system, scope of a peer is fluctuated time to time in a consequence of his dynamic downloading environment. The buffer width of a peer is no longer a constant even for a constant playback rate. Furthermore, different from the discussion in subsection 2.2, such defined buffer width of a peer is no longer equal to his offset lag. We will still denote by $f_p(t)$ the offset curve of peer $p$, and denote by $\xi_p(t)$ the scope curve of this peer. The buffer width curve of this peer now is defined as $W_p(t) = \xi_p(t) - f_p(t)$. It is worthwhile to notice that, in a real system the buffer width $W_p(t) = \xi_p(t) - f_p(t)$ and the offset lag $L_p(t) = s(t) - f_p(t)$ for a peer are different. We will show that the PP scheme in PPLive is more likely based on buffer width instead of offset lag.

## III. Initial Offset Placement in PPLive

### *3.1 Measurement Method and Measurement Platform*

Crawler method general adopted in p2p measurement cannot be applied directly in measure the startup process of PPLive. A host does not advertise his buffer message to other peers at the beginning of startup process. Thus, no any buffer information of a host can be collected by an outside crawler in this silent stage. The silent stage of a host is then defined as the time interval from the time the host joining the system to the time the host advertising his buffer message. A host in this stage is called a silent host. Hence, one must run client software in a host and use tool such as Ethereal to sniffing all exchanged packets. This is a time consuming work and involves a lot of human operations such as restart a channel or selecting a new channel after each experiment. If each experiment takes 5 minutes, only 480 experiments can be made in a full day. The datasets used in this paper are collected at July 3, 11, 12 and 15, 2007 through above mentioned method on a host connecting to a residential broadband network. More than 2,500 experiments are exercised in five runs. At each run, channels (about 500) are selected in turn following the order shown in PPLive user interface window. The silent stage is short (about 5 seconds in general based on our measurements), but it is critical for a peer to establish his initial protocol status.

Once advertises his buffer message to other peers, a host enters a stage named as the advertising stage. Different from the silent stage, the progress status of a peer at the advertising stage can be measured through either monitoring a client or an outside crawler. Both methods have their advantages and drawbacks. For example, detailed and complete data are more easily collected in client monitoring, but loose data are inevitable in outside crawling. In other sides, client monitoring is more time consuming than outside crawling. More important, client monitoring can only find the behavior of a special host in a special network connection, but outside crawling can find generic behavior from variety hosts in different network conditions. In the discussion of this section, we will use data collected from Apr. 2 to Jul. 15,2007 through our crawler to study the advertising stage.

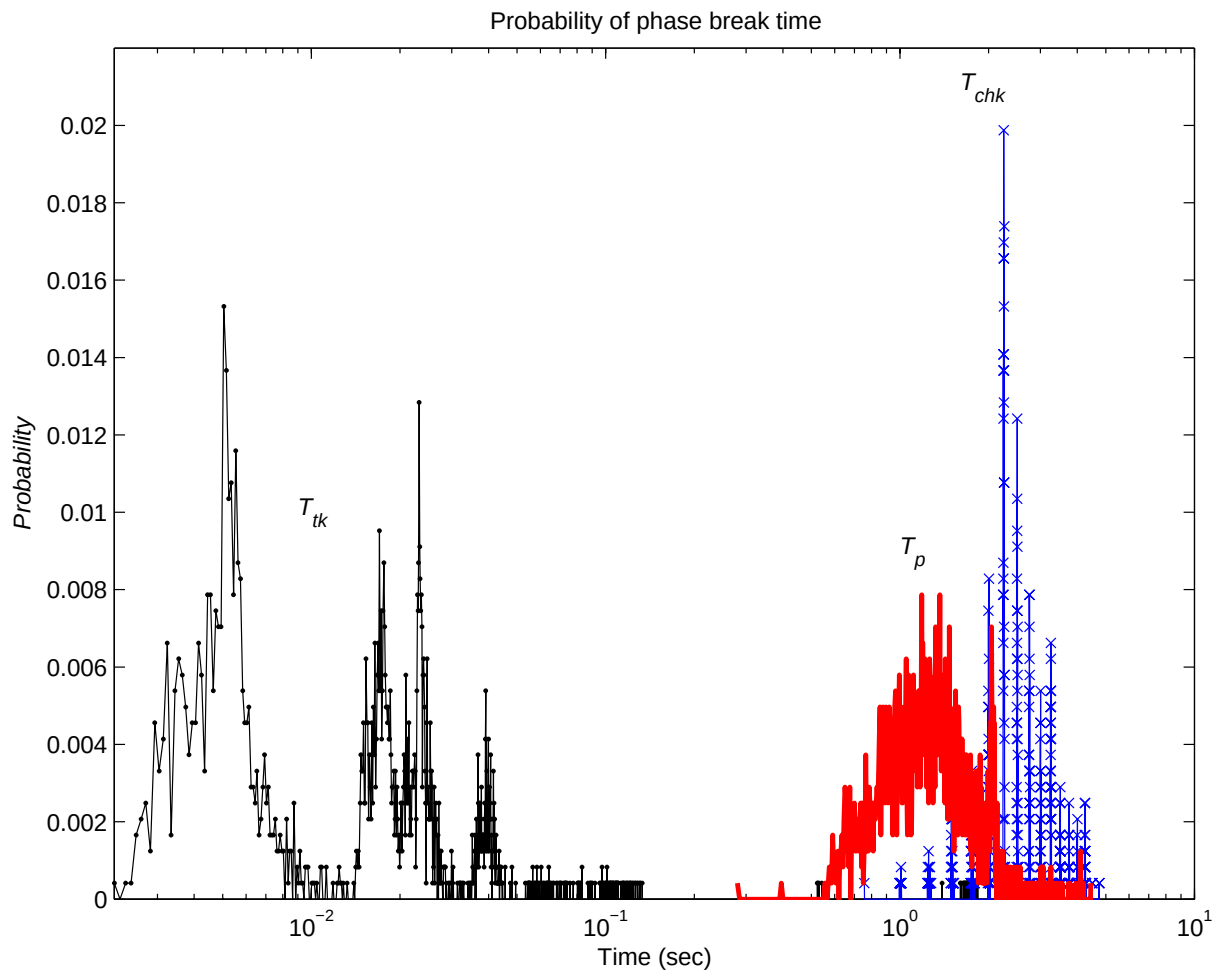

Figure 3. Distributions of the time of different event time

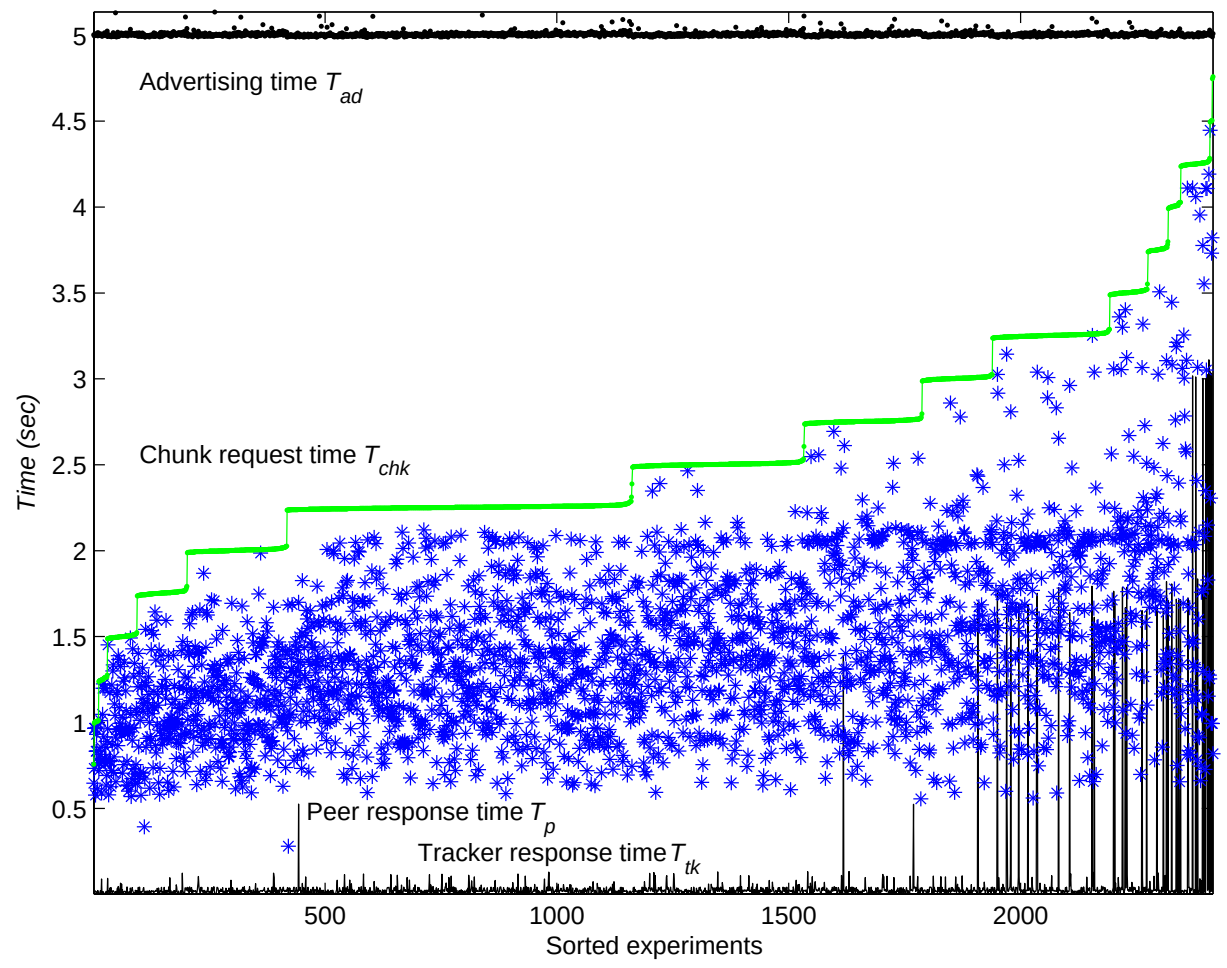

Figure 4. Joint distributions of the time of different events

### *3.2 Event Time Distribution in PPLive*

Events and their time that will be studied in this section are listed in figure 2. All of them except to the *host advertising time* $T_{ad}$ are defined in subsection 2.3. The host advertising time here is defined as the time a host starts to advertise his buffer message to other peers. It breaks the silent stage and the advertising stage of a host. As we have mentioned before, the time of $T_{tk}$, $T_p$, $T_{chk}$, and $T_{ad}$ is measured at the silent stage. The *offset initial time* $T_{off}$ is measured at the advertising stage. Since we use two different platforms to measure the silent and advertising stages, we could not use the time when host registration as the time reference for the offset initial time $T_{off}$. Fortunately, the host advertising time $T_{ad}$ can be measured by both of those two platforms. Hence, we will use $T_{ad}$ as the reference point in measuring $T_{off}$.

There are 2502 experiments in our dataset in silent stage measurement. Ten of them (0.4%) are omitted in our following analysis since in those experiments host advertises his buffer message before receiving any buffer message from other peers [1]. Our data show that the advertising time is a multiple of 5s and most of experiments have $T_{ad}$ ≈5s. Thus, a normal silent stage only lasts for 5 seconds. We find that only 79 (3.2%) experiments have an advertising time larger than 5.2s. Therefore, we will only study those 2413 experiments that satisfy $T_{ad}<T_p$. and $T_{ad}<5.2$s. Probability distributions of phase break time: $T_{tk}$, $T_p$ and $T_{chk}$ are depicted in figure 3. Most trackers return their responses within 0.02 seconds. Peer response time is more like evenly distributed within the time interval from 0.9s to 2s. The mean value for $T_{tk}$, $T_p$ and $T_{chk}$ are 0.058s, 1.419s and 2.566s respectively.

Figure 4 shows the joint distribution of those four time instances. The line in top is the distribution of Advertising time $T_{ad}$. As we have mentioned before, this time is almost constant for all experiments. The curve at the second top is the chunk request time $T_{chk}$. Chunk request time happens at a discrete time instance with a gap of 1/4 second. The blue stars below the chunk request curve are the peer response time $T_p$ and the curve at the bottom is the tracker response time $T_{tk}$. Only extremely large value in tracker response time may introduce a large chunk request time. However, the inverse does not necessarily true. A large chunk request time may follow a small peer response time and/or a tracker response time. No times are tightly dependent.

In the measurement of the advertising stage, our crawler keep to trace each peer in a given channel and records the buffer message returned by each peer with a time stamp to indicate the time this message is received. Thus we will have the offset value of $f_p(t)$ at discrete time instances $\{t_{p,i}\}$ for each peer $p$. Not every peer caught by our crawler is a host since many peers enter the system earlier than our crawler does. A host should have a constant offset at his earlier records and then increased offsets in his later records. There are two problems in inferring the exact value of $T_{off}$ from records of a given host. One is that, the reference point $T_{ad}$ is often missed in the records. A crawler knows a host only through the peer list of the tracker. Hence, in most of the cases, a host already starts his process before our crawler queries him. Another problem is to find the exact time that a host start to drain his buffer because the time span between to consecutive records for some host may last for several tens seconds. Buffer occupations are used to overcome the first problem. The *buffer occupation* $\eta_h(t)$ for a host $h$ is defined as the number of 1s in the buffer map advertised by this host at time $t$. We will only choose those hosts with initial buffer occupations smaller than a certain given threshold, and then use the time stamp of the first record as the estimation of the advertising time $T_{ad}$ of this host. To check if the threshold in initial buffer occupations will introduce biases, different threshold for initial buffer occupations are compared. To overcome the second problem, we use two different methods to estimate the time when a host changes his offset. More precisely, let $t_1$ and $t_2$ be the earliest time pair that our crawler receives two consecutive reports with different offset values $f(t_1)\neq f(t_2)$ from a host. We have tried two different methods to estimate the offset change time $t$. One is based on a simple arithmetic average (AA) $t=(t_1+t_2)/2$. Another is based on linear interpolation (LI) $t=t_2-( f(t_2)-f(t_1) )/r$. The initial offset time $T_{off}$ for a host then is calculated as $T_{off}=t-t_0$, where $t_0$ is the time our crawler receiving the first report from this host. The means and standard deviations of $T_{off}$ estimated by AA and LI with a different initial buffer occupation threshold of 50, 100 and 500 chunks are listed in Table 1. Distributions of $T_{off}$ are also drawn in figure 5. Since all methods give out similar results that validates our methods in estimation of offset setup time. It is worthwhile to notice that, the offset setup time listed in table 1 is inferred from the crawled traces. It is the time interval of $t_{off}-T_{ad}$, only part of the real offset setup time. The real offset setup time is the time interval $t_{off}-T_p=(t_{off}-T_{ad})+(T_{ad}-T_p)\approx 70$s.

[1] This may be introduced by two possible reasons. One is loss data in Ethereal. Lost data in Ethereal are reported by some researchers. The other is that, there may be some peer reported his buffer message but we cannot recognize the message format of this peer, since we do not resolve all message formats in PPLive.

Table 1. Mean and standard deviation of offset setup time

| Threshold of initial buffer occupation | Num. of samples | Arithmetic Average | | Linear Interpolation | |
|---|---|---|---|---|---|
| | | Mean | Sd | Mean | Sd |
| 50 | 854 | 67.04 | 5.49 | 66.99 | 4.84 |
| 100 | 1909 | 66.56 | 5.32 | 66.46 | 5.07 |
| 500 | 3603 | 65.20 | 6.28 | 65.14 | 6.09 |

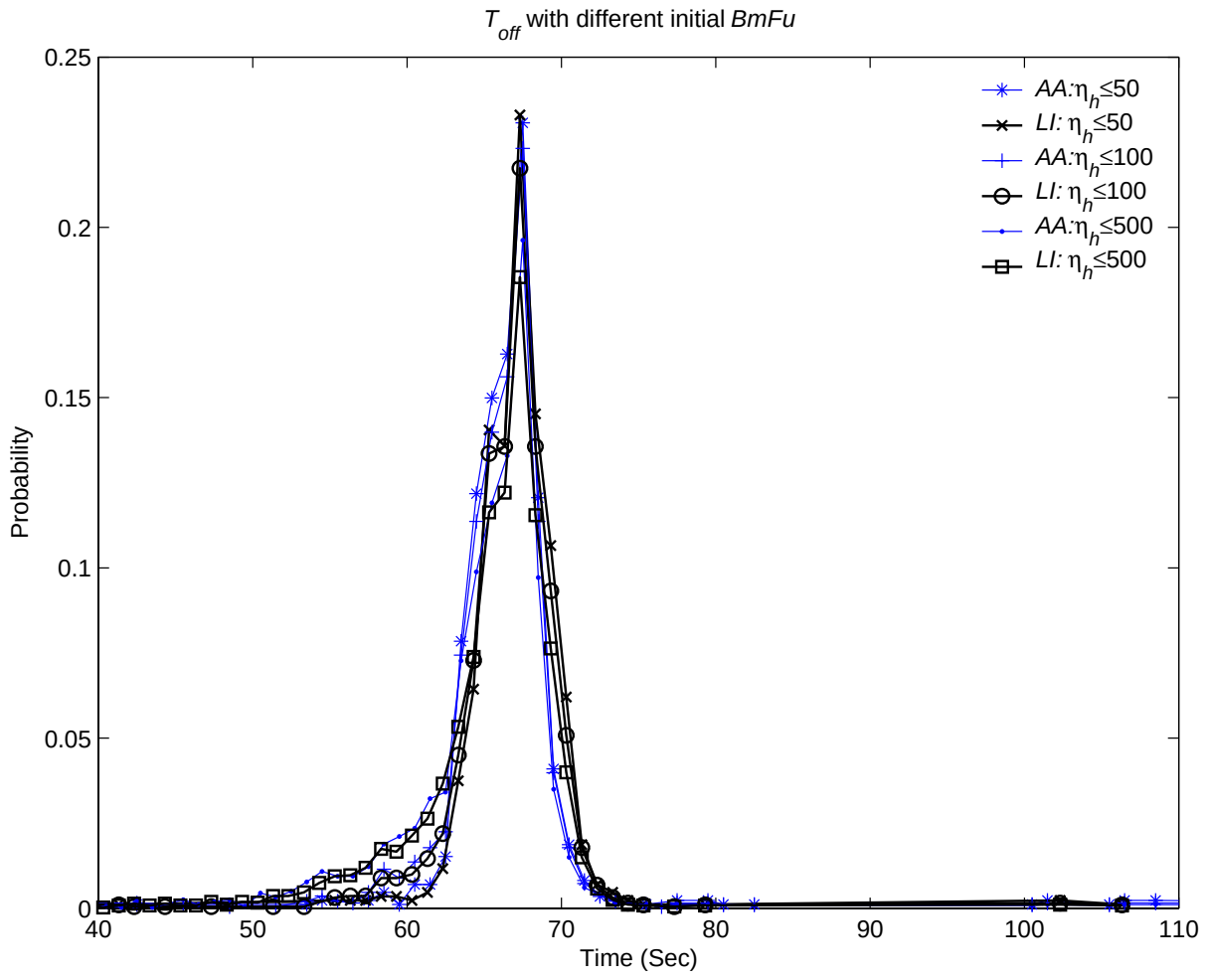

Figure 5. Distributions of offset setup time

### *3.3 Initial Offset Selection*

To check if our simple model for initial offset placement is applied to PPLive, we will try to draw the offset and scope of the first peer along with the offsets placed by host and reported by tracker in one picture (figure 6). Since all related parameters involved are reported at their first time, we will omit the time in their expressions. For example, we will denote by $f_p$ and $\xi_p$ the offset and scope of the first peer reported at the first time respectively. Similarly, we will denote by $s$ and $f_{tk}$ the *TkOffMax* and *TkOffMin* the tracker reported at the first time respectively. Since each experiment falls to a different chunk ID region, for easy visual inspection, we set the zero in y-axis of Figure 6 to represent the $f_{tk}$ for each experiment. The red lines are the tracker buffer width $W_{tk}$. The above one is $W_{tk}$ and below one is $-W_{tk}$. Recall equation (31) that the tracker buffer width is a scaled playback rate in PPLive, one can find that PPLive mainly serves two classes of video playback rate, one is about 10 chunks/s (the right most region in this figure) and the other is about 6 chunks/s (the middle region). The dots marked with *black '.'* and *green 'x'* stand for $\xi_p-f_{tk}$ and $f_p-f_{tk}$ respectively. The distance between *black '.'* and *green 'x'* in same vertical direction is the buffer length $W_p=\xi_p-f_p$ of the first peer in a given experiment. Similarly, The distance between top red line and *green 'x'* in same vertical direction is the offset lag $L_p=s-f_p$ of the first peer. This gives an evidence of variable buffer length scheme adopted by PPLive[2]. It also shows that the tracker has a buffer width much smaller than the buffer width of peers in PPLive. Based on the discussion in section 2, this also indicates that the FP scheme is not adopted by this system. The *blue '*'* in the figure is the initial offset $\theta_h-f_{tk}$ chosen by the host in each experiments. If a host uses a fixed length buffer scheme like the one discussed in [11,12], all *blue '*'* would be aligned in a horizontal line. Actually the *blue '*'* in the figure are distributed with a shape similar to the shape of *green 'x'*. This suggests that the PP scheme in section 2 is very much likely to be adopted by a PPLive host. Since peer buffer width $W_p=\xi_p-f_p$ and peer offset lag $L_p=s-f_p$ are different in PPLive, we will have following two possibilities in implementing our simple model in PPLive. The PP scheme based on the offset lag:

$$\theta_h=f_p+\alpha_L L_p. \quad (32)$$

and the PP scheme based on the buffer width:

$$\theta_h=f_p+\alpha_W W_p. \quad (33)$$

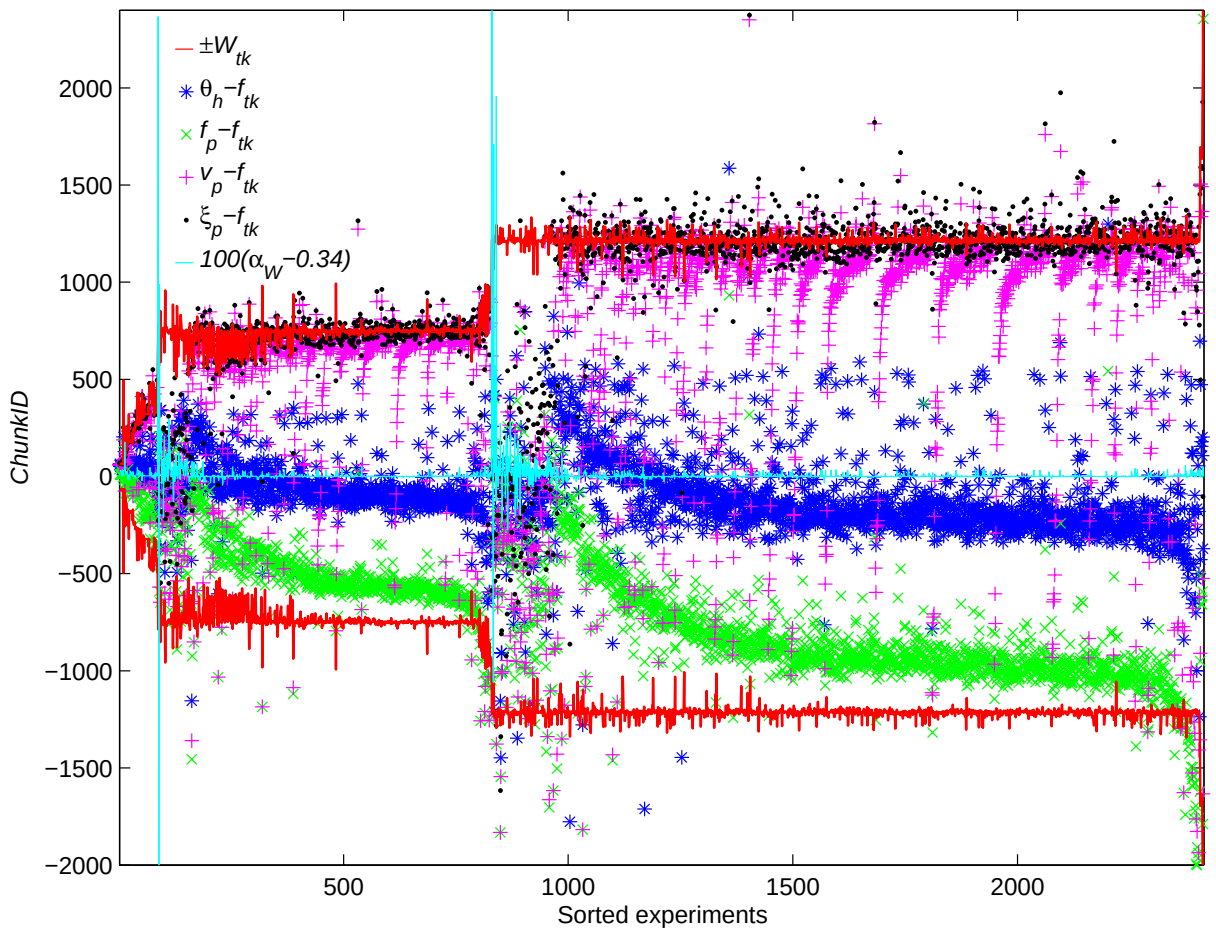

Figure 6 The relations of host initial offset with the buffer status of tracker and the first peer

[2] If we draw the buffer length of all peers in one channel, we can find similar phenomena.

Figure 7 shows the distributions of $\alpha_W=(\theta_h-f_p)/W_p$ and $\alpha_L=(\theta_h-f_p)/L_p$. The PP scheme based on peer buffer width is more likely since $\alpha_W$ has more sharp distribution. The peak of both distributions is at 0.34. This suggests a placement coefficient of 0.34. The cyan line in figure 6 shows the value of $100(\alpha_W-0.34)$, the scaled errors of placement coefficient. One could find that $\alpha_W\approx0.34=\alpha^*$ for most of experiments. Many factors may make the calculated placement coefficient $\alpha_W$ different from $\alpha^*$. Since not all signaling formats used in PPLive are resolved, some peers may use a format we do not know yet to report his buffer message to a host. In this case, the first peer in our analysis may not be the real first peer. Except that, if the first peer is poor, for example too small playable video $V_p$ in the first peer, a host may wait for a better peer. The playable video $V_p$ is defined in [6] as the number of contiguous chunks in the buffer map, beginning from the offset. It is depicted by a mark of *pink '+'* in figure 6.

Intuitional, using a placement based on buffer width is to facilitate the initial chunk fetching. The initial offset based on offset lag may be too large such that no any required chunks are available on all known peers currently. If it is based on buffer width, the host can always find some required chunks from at least one peer. However, if the buffer width of the first peer is very low, the placed host offset will be significantly lagged. Our analysis in section 4 will show that, offset lags may go unbounded when the placement is solely based on the buffer width of the first peer. Certain mechanism to filter out poor first peer is needed to stabilize the performance of the PP scheme based on buffer width.

As shown by the line marked by squares on figure 7, nearly 90% experiments have at least one "best" peer. The placement coefficient based on this "best" peer is about 0.34±10%. This indicates that PPLive does implement a mechanism to select good peer. We do not exactly know what mechanism it is, but a good peer seems to have enough playable video $V_p$ in his buffer. In this paper, a good peer is defined as the earliest

TABLE 2.
PP BASED ON GOOD PEER

```
Good Peer Selection {
  Set timer
  RecPeerResponse( p)
  θ=f_p+α_W W_p
  If (V_p<0.36L_p)
    While (not timeout)
      RecPeerResponse( p)
      If (V_p≥0.36L_p)
        θ=f_p+α_W W_p
        break;
      endIf
    endWhile
  endIf
}
```

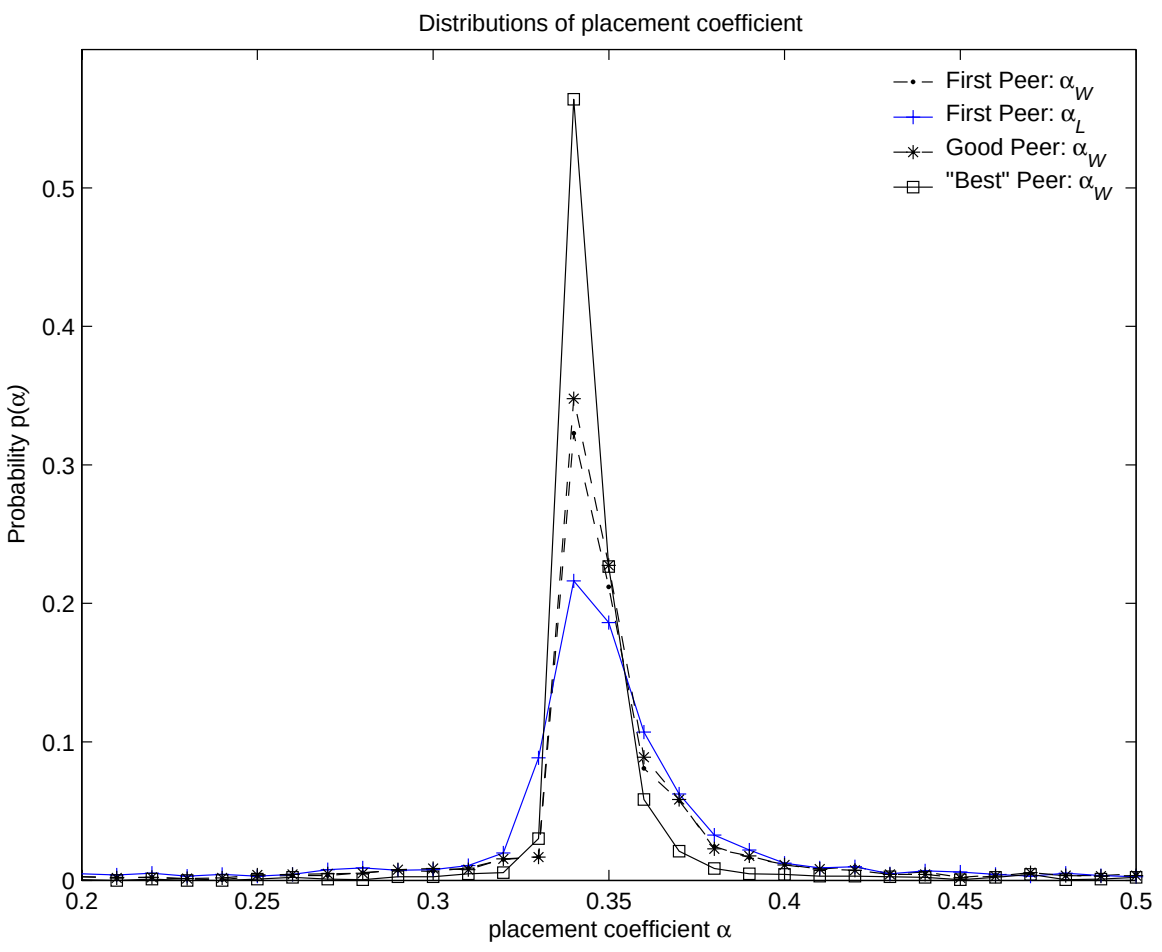

Figure 7. Distributions of placement coefficients

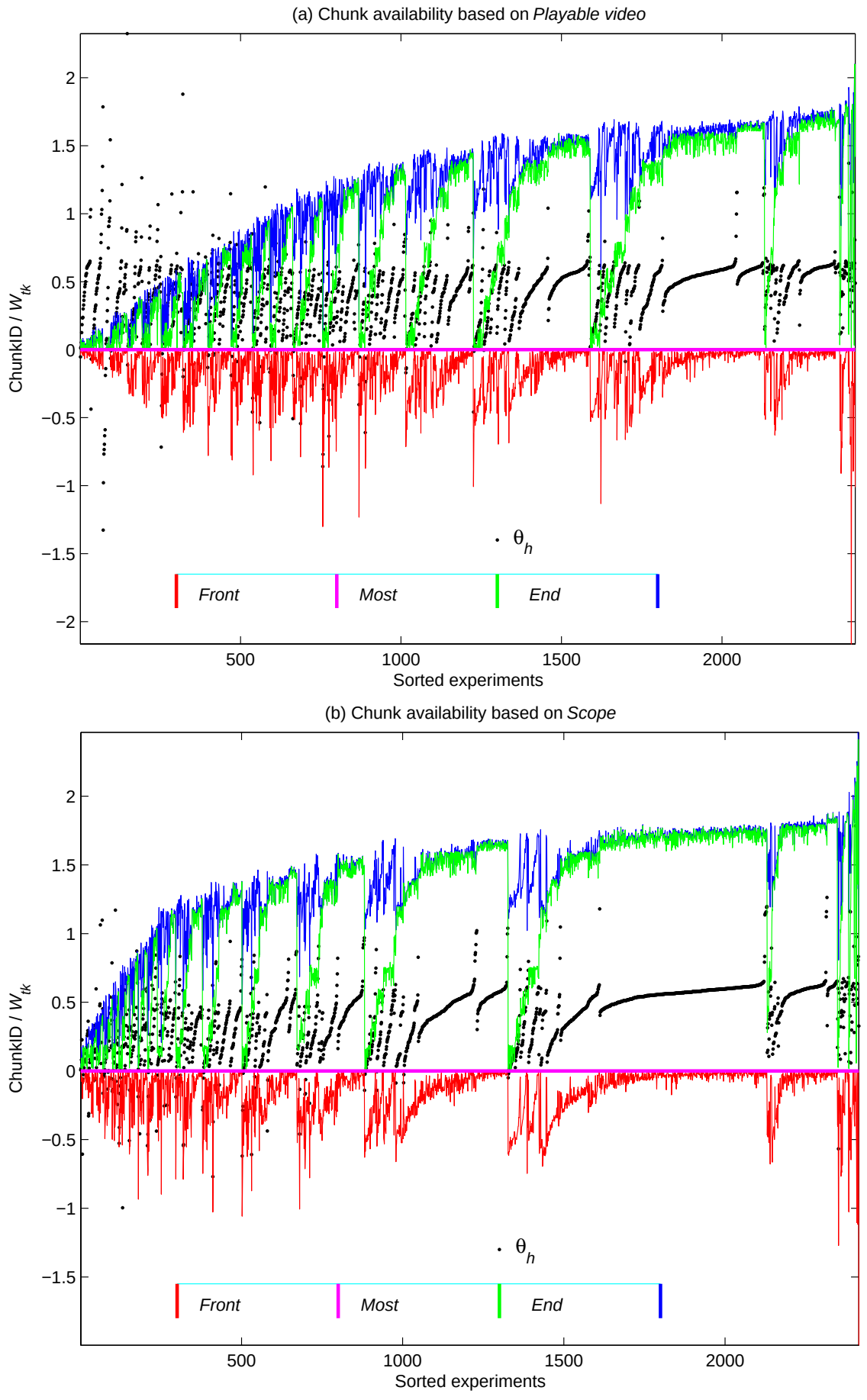

Figure 8. Chunk availabilities

peer with a playable video $V_p \geq 0.36L_p$. Exact protocol used in good peer selection is listed in Table2. The line in astral mark shows the placement coefficient based on the good peer. In our dataset, 71% of experiments have a placement coefficient within the range 0.34±10% based on first peer, it is increased to 76% based on good peer.

### *3.4 Chunk Availability and Initial Offset*

On average, a host receives buffer messages from 4.69 peers before fetching any chunks. The availability of a chunk is defined as the number of known peers who have this chunk. Following two reasons make us to study the availability problem. Firstly, from a view of performance, larger availability around initial offset means good chunk fetching performance. Secondly, from a view of protocol designing, one may ask if the initial offset in PPLive protocol is designed on a current local optimization solely (easier to fetch chunks immediately) or some how balanced with possible buffer distribution of other peers who may not know currently. Instead of to count the number of peers for every chunk, in this paper, we will study chunk availability on *Playable video* and *peer scope* basis respectively. Chunk availability of a chunk on the sense of playable video (peer scope) is the number of peers whose playable video $v_p=f_p+V_p$ (scope $\xi_p$) is larger than the ID of this chunk. The real number of peers who have this chunk is in between of those two availabilities *playable video availability≤ Real availability≤scope availability*.

Figure 8(a) and 8(b) show the playable video and scope availabilities respectively. In those subplots, the chunk segment with largest availability is named as the '*Most*'. If there are more segments that have the largest availability, we will choose the one with largest segment length. The '*Front*' and '*End*' are those two segments they are adjacent to the segment '*Most*'. Chunks in the '*Front*' ('*End*') have Chunk ID smaller (larger) than that of chunks in the '*Most*'. There are 4 points to partition the interval *I* composed by those three consecutive segments *I*=['*Front*', '*Most*', '*End*']. We will name the point with the smallest ID within *Front*, the point with the smallest ID within '*Most*', the point with the largest ID within '*Most*' and the point with the largest ID within '*End*' as the front point (*fp*), low most point (*lmp*), up most point (*ump*) and endpoint (*ep*) respectively. For each experiment, we have an initial offset $\theta$ and four points [*fp*, *lmp*, *ump*, *ep*]. Shifting them to $\theta$–*lmp*, *fp*–*lmp*, *ump*–*lmp* and *ep*–*lmp* aligns a zero *lmp* for all experiments in one subplot. In those subplots, the blue, green and red lines are corresponding to the points of *ep*, *ump* and *fp* respectively. The '*Most*' segment is significantly larger than its two adjacent segments in most experiments. The black dots are the initial offsets. The initial offset of most experiments falls in the section of most available segment in both of playable video and scope senses. Host could fetch chunks around initial offset from more than three neighbor peers in more than 70% of experiments. That indicates a good performance on the placement of initial offset in PPLive. Noticeable black dots can be noted outside the most available region in figure 8 (a) and (b). That may indicate that the initial offset selection in PPLive protocol is not solely based on the current availability.

### *3.5 Summary on the Initial Offset Placement of PPLive*

Through above measurement based study, the PP scheme based on buffer width is very likely adopted by PPLive with a placement coefficient 0.34 and offset setup time about 70 seconds. As we have discussed in section 2, a good peer selection mechanism is also likely adopted by this system. For the stability of this placement in PPLive, we have measured a value of 208.3 for normalized average buffer width $E(W)/r$. With a placement coefficient of 0.34, the offset setup time is about 70.82s for a stable placement. It is very close to the offset setup time we have measured. This implies a stable placement in PPLive.

## IV. Conclusion and Future Work

Initial offset placement in large buffer systems is discussed in this paper. Theoretical models are established and verified by a real system. Peer startup process in a p2p live streaming system is not only the initial offset placement. Initial chunk fetching strategy is also critical in the protocol designing for this peer stage. We will put this in our future researches.